\documentclass[%
 reprint,
 amsmath,amssymb,
 aps,
]{revtex4-2}

\usepackage{graphicx}% Include figure files
\usepackage{dcolumn}% Align table columns on decimal point
\usepackage{bm}% bold math
\usepackage[font=small,labelfont=bf,width=1\linewidth]{caption} 
\usepackage{physics}% extrem nützliches packet für viele symbole was viel in der physik gebraucht wird
\usepackage{xcolor}
\usepackage{mathtools}

\newcommand{\Op}[1]{\hat{\mathrm{#1}}}    %Operator für bessere Übersicht

\usepackage{tcolorbox}
\tcbuselibrary{theorems}

\newtcbtheorem[number within=section]{definition}{Definition}%
{colback=black!7,colframe=white!50!black,fonttitle=\bfseries}{th}

\newtcbtheorem[no counter]{axiom}{Axiom}%
{colback=blue!7,colframe=blue!50!black,fonttitle=\bfseries}{th}

\newtcbtheorem[number within=section]{remark}{Remark}%
{colback=violet!10,colframe=violet!80!black,fonttitle=\bfseries}{th}

\newtcbtheorem[number within=section]{Herleitung}{Derivation}%
{colback=red!7,colframe=red!50!black,fonttitle=\bfseries}{th}
\begin{document}

\preprint{APS/123-QED}

\title{A Didactic Journey from Statistical Physics to Thermodynamics}% Force line breaks with \\
%\thanks{A footnote to the article title}%

\author{Michael Riedl}
\email{michael.riedl@jku.at}
 \affiliation{ Institute for Theoretical Physic - Johannes Kepler Universit\"at, Altenbergerstr. 69, A-4040 Linz, Austria}%Lines break automatically or can be forced with \\
  
\author{Mario Graml}%
\email{mario.graml@jku.at}
\affiliation{%
School of Education - Johannes Kepler Universit\"at, Altenbergerstr. 69, A-4040 Linz, Austria}

\date{\today}% It is always \today, today,
             %  but any date may be explicitly specified

\begin{abstract}
%This work reviewed the foundational concepts of entropy in statistical physics, with an emphasis on its role in quantifying the uncertainty or surprise associated with a system's state. The discussion is focused on mixed states derived from quantum mechanics. The aim is to provide a comprehensive yet accessible guide, transitioning from an understanding of entropy to the development of phenomenological thermodynamics, and elucidating key terminology along the way.
%This paper presents a focused exploration of entropy's fundamental properties, elucidating its significance in statistical physics. It begins with a single axiom and derives the Maxwell relations, stability conditions, and establishes the concavity of entropy, dispelling misconceptions surrounding Lagrange parameters' physical interpretations. Motivated by this, it delves into the geometric, mathematical, and experimental underpinnings of the Legendre Transformation, culminating in a comprehensive formalisation. By means of a rigorous analysis, we define dependencies and potentials along with their natural variables, elucidating their role in statistical physics. Finally, we derive thermodynamic axioms from the principles of statistical physics, thus bridging theoretical frameworks and offering deeper insights into the foundational concepts of thermodynamics.
This paper offers a pedestrian guide from the fundamental properties of entropy to the axioms of thermodynamics, which are a consequence of the axiom of statistical physics. It also dismantles flawed concepts, such as assigning physical meaning to Lagrange multipliers and numerous others. This work also provides a comprehensive understanding of the Legendre transform via geometrical, mathematical and physical insights, as well as its connection to the experimental setup. The central result of this paper is the comprehensive formalisation of key concepts, including ensembles, variable dependencies, potentials and natural variables. Furthermore, the framework of thermodynamics, the state function and the Euler inequality are rigorously proven from the axiom of statistical physics.
\end{abstract}

%\keywords{Suggested keywords}%Use showkeys class option if keyword
                              %display desired
\maketitle

%\tableofcontents
%\section{Paperaufbau}
%Alles vorbereitet, damit wir loslegen können. \cmg{Kommentare Graml} \\ \cmg{Befor einreichen müssen wir die Boxen entfernen}
%\cmk{Abhängigkeiten besser rausschreiben, Introduction besser verkaufen/Reservoirs mit Legendre Physikalisch verkaufen, Unterschied  U E ausarbeiten, Thermodynamisch mit bekannten verbinden, intensive-extensive???}\cmr{Reservoire, experimentel interpretieren (isolierung, wie entkoppelt man experimentel, U,F,G thermo dazuschreiben}\cmg{Conclusion klare unterscheidung von Konzepten besser verkaufen. Wir können unterschiede niederschreiben, systemgrenzen etc, Abstract aufgeilen.}
\section{Introduction}
In the framework of statistical physics, physical phenomena are only expressible through probabilities \cite{Callen1991,neumaier2011classical}. Central to this framework is the concept of entropy \cite{LANDAU19801,MackeWilhelm1963ELdt,FrederikStat,Fultz2020,Khomskii2010}, a fundamental quantity that encapsulates the average amount of missing information or surprise associated with a system's state. Through the lens of quantum mechanics, these probabilities are encoded in what is known as a mixed state, which is a mixture of pure states, each weighted by their respective probabilities \cite{LANDAU198079,FrederikVO,Neumann}. This work aims to give a pedestrian guide as mathematically precise as needed to develop the framework starting from entropy, and ending in phenomenological thermodynamics. It allows us to define precisely the heavily misused terms in the context.\\   
In sec. \ref{sec:idea} we present the basic ideas of entropy, leading to the definition of entropy, and then demonstrate all its properties in its special tailor-made example.
Entropy is determined in sec. \ref{sec: Determining} with the mathematical formulation of the physical intuition on the given experimental data. A comprehensive discussion of the connection between theory and experiment is given in sec. \ref{sec: Connection} and is concluded with the stability conditions and Maxwell relations.
Changing the tunable parameters in the experimental setup must be incorporated into the theory, as discussed in section sec. \ref{sec: Legend}. This will be done threw a full exploration of the Legendre transformation in terms of its geometrical meaning, mathematical formulation and physical consequences for the experimental setup.
The principal findings, as detailed in sec. \ref{sec:form}, are the precise formulation and definition of ensembles, their representation, potentials and natural variables. 
Finally, in sec. \ref{sec: Thermo}, the framework of phenomenological thermodynamics, including the state function and Euler inequality, is derived from the axiom of statistical physics.
\section{Idea and Definition of Entropy} \label{sec:idea}
In statistical physics, physical concepts are described using statistics and probabilities. In the framework of quantum mechanics, these probabilities are encoded in the eigenvalues of the mixed state $\Op{\rho}$ (hereafter referred to as the state). The state consists of pure states $\Op{\rho}_i$ with associated probabilities $p_i$. The objective is to approximate the state by mathematically determining the $p_i$ as accurately as possible. The approximate state is quantified by a number that satisfies the following three conditions: a value of zero represents complete knowledge of the state, while the maximum value represents complete uncertainty. When quantifying independent systems together, the numbers of the source systems should add up.
The operator $\ln(\Op{\rho})$ fulfils these three properties of quantifying the state of a system. The resulting functional $S[\Op{\rho}]$ is known as \textit{von Neumann}-\textbf{entropy}.
\begin{definition}{Entropy}{}
    The entropy $S[\Op{\rho}]$ for a given state $\Op{\rho}$ is defined as
    \begin{align}
        S[\Op{\rho}] \coloneqq-k \Tr\left(\Op{\rho} \, \ln(\Op{\rho})\right ) \label{def_S}
    \end{align}
    where $k>0$ is a proportionality constant.
\end{definition}
Entropy is a measure of the average missing information. It can also be interpreted as a measure of how surprised someone would be at the result of an experiment. The following examples will clarify the intuitions regarding surprise and average missing information.\\ \\
   \textbf{Complete knowledge}: If the state is fully known, the inner product $\braket{\phi_i}{\Op{\rho}_j(\phi_i)}=\delta_{ij}$ leads to
    \begin{equation}
        S[\Op{\rho}]=-k\Tr\left(\Op{\rho}\ln\left(\Op{\rho}\right) \right)=-k\sum\limits_i \delta_{ij}\ln\left(\delta_{ij}\right)=0 \label{S_pure}
    \end{equation}
  Eq. \eqref{S_pure} illustrates that there is no surprise, if the experiment always determines the pure state $\Op{\rho}_j$.\\
    \\
    \textbf{No knowledge}:
    On the other hand, to obtain a maximally surprising result from the experiment, no prior knowledge about the state is assumed. This implies that the state is consisting of uniformly distributed pure states. Assuming that the state consists of $n$ such pure states, then
    \begin{equation}
     S[\Op{\rho}]=-k\Tr\left(\Op{\rho}\ln\left(\Op{\rho}\right) \right)=-k \sum\limits_{j=1}^n \frac{1}{n}\ln\left(\frac{1}{n}\right)=k\ln(n)  \label{S_max} 
    \end{equation}
  Eq. \eqref{S_max} represents the maximum value of $S$ and is known as the \textbf{Boltzmann entropy}, with $n$ being the number of uniformly distributed pure states.  
  The proof of the third property is presented in the appendix \ref{A.S_add}, where it is shown that for two \textbf{independent} systems A and B, the entropy is additive:
  \begin{equation}
      S[\Op{\rho}_{\text {AB}}]=S[\Op{\rho}_{\text A}]+S[\Op{\rho}_{\text B}]\label{s_add}
  \end{equation}
  To gain an understanding of the three properties and how to handle them, let us consider an example:\\ \\
\textbf{Three coin tosses}\\ 
In this thought experiment, we will toss a coin three times.     There are only two possible outcomes for each toss: Heads (H) or Tails (T). There are $n=2^3$ different outcomes for tossing a coin three times.
  %\begin{center}
  \begin{figure}
  \captionsetup{width=.935\linewidth ,justification=raggedright, singlelinecheck=false}
       \includegraphics[width=0.9\linewidth]{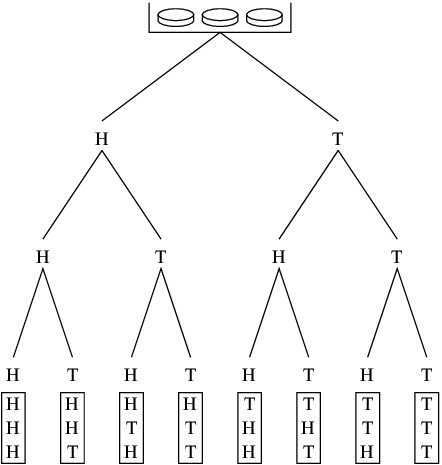}
        \captionof{figure}{Three coin tosses with the single outcomes Heads (H) or Tails (T) is represented. The probability for the single outcome H is given by $p_{\text H}$ and the probability for the single outcome T is given by $p_{\text T}$ respective. The three tosses resulting in $2^3$ different possibilities for the pure state, denoted as HHH, HHT, \ldots, TTT. }\label{fig_CoinToss}
\end{figure}
 % \end{center}
This experimental setup is illustrated in fig. \ref{fig_CoinToss}.
In this setup, we choose $k=\frac{1}{\ln 2}$, so the entropy will be in units of $\log_2$, also known as bits of information. We will consider three different scenarios:\\ \\
\underline{Scenario one involves three fair coin tosses:}\\
Assuming no prior knowledge of the coin, we can infer that the coin is fair, with an \textbf{equally distribution} of H and T. In order to ascertain the state, it is necessary to answer \textbf{three} questions of either H or T. As all pure states are \textbf{equally distributed}, eq. \eqref{S_max} of the Boltzmann entropy can be used to calculate the average missing information, denoted as the entropy:
\begin{equation}
       S[\Op{\rho}]=\log_2(n)=\log_2\left(2^3\right)=3  \label{S_3}
    \end{equation}
In eq. \eqref{S_3}, the average missing information, expressed in units of bits, is \textbf{three}, corresponding to the three answers of H or T.\\
    \\
    \underline{Scenario two consists of three highly skewed coin tosses:}\\
    In this instance, we assume that the coin is biased, with a probability of H, $p_\text H=0.95$. Therefore, we now have more information about the system than we did previously. Since the pure states are not equally distributed, we can no longer use Boltzmann's entropy. However, the state still consists of three independent subsystems, the three coin flips, so we can still use the third property eq. \eqref{s_add} and add up the subsystems:
     \begin{align}
    S[\Op{\rho}_{AAA}]&=S[\Op{\rho}_{A}]+S[\Op{\rho}_{A}]+S[\Op{\rho}_{A}]= 3S[\Op{\rho}_{A}]\notag \\ &=-3\left(p_{\text H}\log_2\left(p_{\text H}\right)+p_{\text T}\log_2\left(p_{\text T}\right)\right) \approx 0.86
    \end{align}%0.859191
The average missing information is 0.86 bits, and one is not so surprised because one can expect two to three H with a probability of $>99$\%.\\ \\ 
    \underline{Scenario three uses a memory coin:}\\
The difference between this setup and the previous two is that the tosses are \textbf{dependent on each other}. The additional rule for this experiment is that the probability of getting the same result as the previous flip is halved. Starting with a fair toss, the probabilities of the pure states are calculated as follows
\begin{itemize}
    \item $p(H,H,H)=p(T,T,T)=\frac{1}{2} \cdot \frac{1}{4} \cdot \frac{1}{8}=\frac{1}{64}$
    \item $p(H,H,T)=p(T,T,H)=\frac{1}{2} \cdot \frac{1}{4} \cdot \frac{7}{8}=\frac{7}{64}$
    \item $p(T,H,H)=p(H,T,T)=\frac{1}{2} \cdot \frac{3}{4} \cdot \frac{3}{8}=\frac{9}{64}$
    \item $p(T,H,T)=p(H,T,H)=\frac{1}{2} \cdot \frac{3}{4} \cdot \frac{5}{8}=\frac{15}{64}$
\end{itemize}
    To calculate the entropy for \textbf{dependent} systems we have to use the \textbf{general formula} eq. \eqref{def_S}:
    \begin{align}
        S[\Op{\rho}]=-\Tr \left(\Op{\rho}\log_2\left (\Op{\rho} \right) \right)=-\sum\limits_{i=1}^8 p_i\log_2(p_i) \approx 2.7 <3 \label{S_mem}
    \end{align}
As can be seen in eq. \eqref{S_mem}, one is still surprised, but due to the information about the switching probabilities, the surprise is not as great as the initial value of 3.
\section{Determining the Entropy from Experimental Data}\label{sec: Determining} 
As we have seen, entropy allows us to quantify our knowledge of a state. In practice, however, we cannot determine the distribution of the state and can only measure certain quantities of the system. From these experimental results, we want to extract the probability distribution of the state in the most unbiased way. This leads to the axiom of statistical physics:
\begin{axiom}{Axiom of Statistical Physics}{}
    If we cannot, or do not want to, know the state of a system completely, then we have to construct the probability distribution of the state, such that the state
        \begin{itemize}
            \item[1.)] reproduces the experimental measurements  
            \item[2.)] is maximally unbiased.
        \end{itemize}
\end{axiom}
This axiom can now be equivalently stated: 
\begin{axiom}{Equivalent Formulation}{}\label{Ax}
    Let $\Op{\rho}$ be the unknown state of the system, we want to find $\Op{\rho}$ such that
        \begin{itemize}
            \item[1.)] $\expval{\Op{R}_i}_{\Op{\rho}}=r_i  \Leftrightarrow \Tr\left(\Op{\rho} \Op{R}_i\right)=r_i $ and
            \item[2.)] $ S[\Op{\rho}] $ is maximal. 
        \end{itemize}
    where $\Op{R}_i$ denotes the $i^{\text{th}}$ observable, $r_i \in \mathbb{R}$ is the experimental measurement of $\Op{R}_i$.
\end{axiom} The first part of the axiom ensures that the constructed state reflects the measured values $r_i$. The second part of the axiom ensures maximum ignorance of the unknown information, resulting in the maximization of the entropy.\\ 
This consideration leads to a variation of $S[\Op{\rho}]$ with boundary conditions. We solve this problem using the method of Lagrange multipliers. The first part of the axiom gives us the number of Lagrange multipliers $k\lambda'$. Additionally, the property of the state $\Tr(\Op{\rho})=1$ provides us with an additional Lagrange multiplier $k\Omega'$:
   \begin{subequations}
    \begin{align}
        S[\Op{\rho}]=-k \Tr\left(\Op{\rho} \ln(\Op{\rho})\right)+&k\Omega'\left(\Tr(\Op{\rho})-1\right) \notag \\
        -\sum\limits_{j=1}^J&k \lambda_j^\prime \bigg(\Tr(\Op{R}_j \Op{\rho})-r_j \bigg)\label{S_start}
    \end{align}
    For practical reasons, we scale the multipliers with $k$ so that the variation is independent of $k$. Carrying out the variation of eq. \eqref{S_start} and setting it to zero, as required by the method
    \begin{equation}
        \forall \Op{\tau}: \hspace*{5pt} 0\stackrel{!}{=}\delta_\tau S[\Op{\rho}] \coloneqq\lim_{\epsilon \rightarrow 0} \frac{S[\Op{\rho}+\epsilon \Op{\tau}]-S[\Op{\rho}]}{\epsilon}
        \label{var_prinz}
    \end{equation} 
    leads to the condition
    \begin{equation}
      0=-k \Tr\left(\Op{\tau} \Bigg(\ln\left(\Op{\rho}\right) -\Omega + \sum_{j=1}^J \lambda_j^\prime \Op{R}_j \Bigg)\right) \label{cond} 
    \end{equation}
    with $\Omega=\Omega'-1$.
    In accordance with eq. \eqref{var_prinz}, eq. \eqref{cond} must hold for all $\Op{\tau}$, leading to a more refined condition:
    \begin{equation}
        \ln \left(\Op{\rho}\right) -\Omega + \sum_{j=1}^J \lambda_j^\prime \Op{R}_j =0 \label{cond2}
    \end{equation}
    From the resultant expression in eq. \eqref{cond2}, we derive an equation for $\Op{\rho}$:
    \begin{equation}
        \Op{\rho}(\Omega,\lambda^\prime)=\exp\left(\Omega-\sum_{j=1}^J \lambda_j^\prime \Op{R}_j\right) \label{rho_ex}
    \end{equation}
    where we introduced the shorthand $\lambda^\prime=\left(\lambda_1^\prime,\lambda_2^\prime,\dots,\lambda_J^\prime\right)$. The final step to complete the variation principle is to determine the Lagrange multipliers $\lambda'$ and $\Omega$. For this purpose, we utilize our boundary conditions. Using the condition $\Tr(\Op{\rho})=1$ fixes $\Omega$ to
    \begin{equation}
         \Omega(\lambda^\prime)=-\ln\left(\Tr\Bigg(\exp\bigg(-\sum_{j=1}^J \lambda_j^\prime \Op{R}_j\bigg)\Bigg)\right) \label{omega_ex}
    \end{equation}
    To derive the remaining $J$ equations, we leverage the property of the method of the Lagrangian parameters:
    \begin{equation}
        \pdv{S(\lambda',\Omega)}{\lambda'_i}=0 \label{cond3}
    \end{equation}
    Calculating the entropy with our derived $\Op{\rho}$ in eq. \eqref{rho_ex} yields
    \begin{equation}
        S(\lambda')=-k\Omega(\lambda')+k\sum\limits_{j=1}^J\lambda_j'r_j
    \end{equation}
      This results in $J$ equations, compactly written as 
    \begin{equation}
        \frac{\partial \Omega(\lambda^\prime) }{\partial \lambda_i^\prime}=r_i \label{r_i_Lagrange}
    \end{equation}
    This system of equations needs to be solved to find $\lambda^\prime(r)$, where $r=(r_1,r_2,\dots,r_J)$. 
       \end{subequations}
    Substituting these expressions into eqs. \eqref{rho_ex} and \eqref{omega_ex}, we obtain the following results:
    \begin{equation}
\Op{\rho}\left(\lambda'\left(r\right)\right)=\exp\left(\Omega(\lambda^\prime(r))-\sum_{j=1}^J \lambda^\prime_j(r) \Op{R}_j\right) \label{rho_fertig}
    \end{equation}
    and
    \begin{equation}
        \Omega(\lambda^\prime(r))=-\ln\left(\Tr\Bigg(\exp\bigg(-\sum_{j=1}^J \lambda_j^\prime(r) \Op{R}_j\bigg)\Bigg)\right) \label{Omeiga_fertig}
    \end{equation}
   We have constructed our state as given in eq. \eqref{rho_fertig} with eq. \eqref{Omeiga_fertig} to ensure that the resulting state $\Op{\rho}$ satisfies the axiom of statistical physics. The associated entropy is given by
   \begin{equation}
       S(r)\equiv S(\lambda'(r))=-k\Omega\left(\lambda^\prime\left(r\right)\right)+k\sum_{j=1}^J \lambda^\prime_j\left(r\right)r_j \label{S_final}
   \end{equation}
    Ultimately, our focus is solely on the relationship between entropy and the measurement. To complete the Lagrangian multiplier method, we must disregard the intermediate dependencies $\lambda'(r)$ outlined in equation \eqref{S_final}. 
\section{Connecting Theory and Experiment} \label{sec: Connection}
This is where a crucial question arises: How does the system respond to changes in the measurements? In the experiment we can measure the changes of $S$ on the measurements $r_i$. These are important quantities, so we define this change as $\lambda_i$ in units of $k$:
\begin{equation}
     k \lambda_i(r) \coloneqq\pdv{S(r)}{r_i} \label{S_der_ex}
\end{equation}
    The connection between $\lambda_i$ and $r_i$ is so fundamental that it gets its own name. The $k\lambda_i$ and $r_i$ are called \textbf{entropic conjugated variables}. To connect the experimental value eq. \eqref{S_der_ex} with our model, we calculate the derivative of eq. \eqref{S_final}:
    \begin{align}
        \frac{\partial S(r)}{\partial r_i} &=-k \sum_{j=1}^J \underbrace{\frac{\partial \Omega(\lambda^\prime)}{\partial \lambda_j^\prime}}_{r_j} \frac{\partial \lambda_j^\prime(r)}{\partial r_i}+k\sum_{j=1}^J \frac{ \partial \lambda_j^\prime(r) }{\partial r_i}  r_j + k \lambda_i^\prime(r) \notag \\&= k \lambda_i^\prime(r) \label{S_der_mod}
    \end{align}	
    The combination of the experimental (exp) values eq. \eqref{S_der_ex} with the model (mod) values eq. \eqref{S_der_mod} yields
    \begin{equation}
         \lambda_i(r)\stackrel{\text{exp}}{=}\frac{1}{k}\frac{\partial S(r)}{\partial r_i}\stackrel{\text{mod}}{=} \lambda_i^\prime(r) \label{exp_mod_comb}
    \end{equation}
    Eq. \eqref{exp_mod_comb} connects the experimental entropic conjugated variables $k\lambda_i$ with our purely mathematical Lagrange multipliers $k\lambda_i'$. It should be emphasised that they are only equivalent if we choose the Lagrangian multipliers as we did in eq. \eqref{S_start}.\\ 
    With the connection to the experiment just made, we can check the meaning of the maximum condition for the entropy for concrete systems. Since the proof gives insight into the properties of entropy other than that it is maximal, we will do it in detail.\\
    To prove the maximum, one needs to compute the Hessian matrix, a matrix of second derivatives, and check whether it is negative definite at the measurements $r$. 
    The Hessian matrix is calculated as:
\begin{equation}
\frac{\partial^2 S(r)}{\partial r_i \partial r_j} = k\frac{\partial \lambda_i (r)}{ \partial r_j} \label{hessian_s}
\end{equation}
Despite the inability to perform the derivative of $\lambda_i$ with respect to $r_j$ abstractly, we can employ the identity:
\begin{equation}
\pdv{\lambda_i}{r_j}=\frac{1}{\pdv{r_j}{\lambda_i}} \label{part_der_iden}
\end{equation}
With eq. \eqref{r_i_Lagrange}, we can rewrite the derivative as:
\begin{equation}
\frac{\partial r_j(\lambda)}{\partial\lambda_i}=\frac{\partial^2 \Omega(\lambda)}{\partial \lambda_i \partial \lambda_j}\label{hessian_omeg}
\end{equation}
As one can see, the eigenvalues of the Hessian matrix of $\Omega$ in eq. \eqref{hessian_omeg} are the inverse eigenvalues of our original Hessian matrix of $S$ in eq. \eqref{hessian_s}. This implies that it is equivalent to showing that the Hessian of $\Omega(\lambda)$ is negative definite. Calculating the Hessian of $\Omega(\lambda)$, we find:
\begin{equation}
    \frac{\partial^2 \Omega(\lambda)}{\partial \lambda_i \partial \lambda_j}=\underbrace{\Tr\left(\Op{R}_i\Op{\rho}(\lambda)\right)}_{r_i(\lambda)} \Tr\left(\Op{R}_j\Op{\rho}(\lambda)\right) - \Tr\left(\Op{R}_i \Op{R}_j \Op{\rho}(\lambda)\right) \label{mat_omeg}
\end{equation}
The matrix elements in eq. \eqref{mat_omeg} exactly represent the negative \textbf{covariance}, denoted \textbf{CoV}, defined as
\begin{equation}
   \!\! \text{CoV}_{\Op{\rho}}(\Op{R}_i,\Op{R}_j)(\lambda) \coloneqq \Tr\bigg(\Op{R}_i \Op{R}_j \Op{\rho}(\lambda)\bigg)-r_i(\lambda) r_j(\lambda) \label{Cov}
\end{equation}
    Since the matrix elements in eq. \eqref{mat_omeg} have the form of the covariance as shown in eq. \eqref{Cov}, it is a standard procedure to prove that the associated matrix is negative definite. We will go through this proof for two reasons: Firstly, it is constructive, providing insight into $\Omega(\lambda)$. Secondly, we can extract relevant information regarding the experiment.\\ 
    \begin{subequations}
    The Hessian of $\Omega(\lambda)$ is negative definite, i.e.
    \begin{equation}
        \forall a\in \mathbb{R}^J: \sum\limits_{j,k=1}^{J} a_j \frac{\partial^2 \Omega(\lambda)}{\partial \lambda_j \partial \lambda_k}  a_k < 0 \label{negative_def}
    \end{equation}
    The proof of eq. \eqref{negative_def} begins by inserting the definition of the covariance in eq. \eqref{Cov}:
\begin{equation}
    \sum\limits_{j,k=1}^J\!\! a_j \frac{\partial^2 \Omega(\lambda)}{\partial \lambda_j \partial \lambda_k}  a_k=\!\!\sum\limits_{j,k=1}^J\!\! a_j\! \left (\expval{\Op{R}_j}_{\Op{\rho}}\expval{\Op{R}_k}_{\Op{\rho}}-\expval{\Op{R}_j\Op{R}_k}_{\Op{\rho}} \right)\!  a_k
\end{equation}
    Next, we use the linearity of the expectation value
    \begin{equation}
    =\expval{\sum \limits_{j=1}^Ja_j\Op{R}_j}_{\Op{\rho}}  \expval{\sum \limits_{k=1}^Ja_k\Op{R}_k}_{\Op{\rho}}- \expval{\sum \limits_{j,k=1}^Ja_ja_k\Op{R}_j\Op{R}_k}_{\Op{\rho}} 
    \end{equation}
    Rename the sum from $j$ to $k$ in the first term  to arrive at
    \begin{equation}
=\expval{\sum\limits_{k=1}^Ja_k\Op{R}_k}_{\Op{\rho}}^2-\expval{\sum\limits_{j,k=1}^Ja_ja_k \Op{R}_j\Op{R}_k}_{\Op{\rho}} \label{before_var}
    \end{equation}
    Upon examining eq. \eqref{before_var}, one can identify the variance, indicated as $\mathbb{V}$:
\begin{equation}
    \expval{\sum \limits_{k=1}^Ja_k \Op{R}_k}_{\Op{\rho}}^2-\expval{\sum\limits_{j,k=1}^Ja_ja_k \Op{R}_j\Op{R}_k}_{\Op{\rho}}=-\mathbb{V}_{\Op{\rho}} \left ( \sum\limits_{k=1}^Ja_k \Op{R}_k \right) \label{after_var}
\end{equation}
The variance is positive, which means that eq. \eqref{after_var} is always less than zero for any $a$, which is exactly what we wanted.\\
    \end{subequations}
    We have just proved that the Hessian of $\Omega(\lambda)$ is negative definite independent of $\lambda$, which implies that the entropy $S(r)$ has the same properties independent of $r$. From a mathematical point of view, we can directly derive the so-called \textbf{stability conditions} for the experiment. We know that the $i^{\text{th}}$ diagonal element of the Hessian eq. \eqref{mat_omeg} are the negative variance of $ \Op{R}_i$ . From that, we obtain the \textbf{one-parameter stability condition} as
    \begin{equation}
        \pdv{r_i}{\lambda_i}<0 \label{stable_1}
    \end{equation}
    For the experiment eq. \eqref{stable_1} tells us that changing one parameter $\lambda_i$ must change the associated $r_i$ with a negative slope, like a self-stabilising system. From mathematics we know that if a matrix is negative definite, then the determinants of all submatrices must also be negative. For changing two parameters in the experiment we can then expect the \textbf{two-parameter stability conditions}:
    \begin{align}
       \det( \begin{matrix}
            \pdv{r_i}{\lambda_i} & \pdv{r_j}{\lambda_i}\\
            \pdv{r_i}{\lambda_j} & \pdv{r_j}{\lambda_j}
        \end{matrix})<0 \Rightarrow
        \pdv{r_i}{\lambda_i}\pdv{r_j}{\lambda_j}<\pdv{r_i}{\lambda_j}\pdv{r_j}{\lambda_i}
    \end{align}
    This principle can be extended to include all $J$ parameters in the experiment.\\
    It should be noted that measuring the change $\pdv{r_j}{\lambda_i}$ may be experimentally difficult or even impossible. However, it can often be determined by using eq. \eqref{hessian_omeg}:
    \begin{equation}
        \frac{\partial r_j(\lambda)}{\partial\lambda_i}=\frac{\partial^2 \Omega(\lambda)}{\partial \lambda_i \partial \lambda_j}=\pdv{r_i(\lambda)}{\lambda_j} \label{Maxwel}
    \end{equation}
    The order of the partial derivatives is of no concern, so it is easy to construct the so-called \textbf{Maxwell-relations} eq. \eqref{Maxwel} which are heavily used in experiments.\\ 
    Above we proved that the Hessian of $S(r)$ is negative for every $r$, meaning that the entropy is even \textbf{concave}. This is a powerful property and will be used to extend the possibilities of modeling the experiments by including \textbf{reservoirs}. The resulting technique we need is the \textbf{Legendre transformation}.
\section{Legendre Transformation} \label{sec: Legend} 
\begin{figure}
  \captionsetup{width=0.935\linewidth ,justification=raggedright}
       \includegraphics[width=0.9\linewidth]{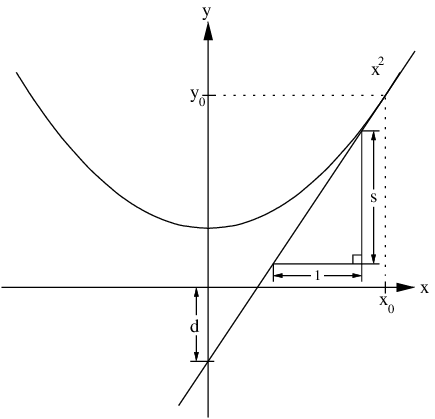}
        \captionof{figure}{The function $y=f(x)$ is plotted based on $x$. An arbitrary point $x_0$ is selected. The slope $s$ of the function $f$ at point $x_0$ is given as $s=f^\prime(x_0)$ and the offset $d$ of the resulting tangent $y=s\, x+d$ at point $(x_0,y_0)$ is given by $d=y- s \, x$.} \label{fig_Lt}
  \end{figure}

\begin{figure}
    \includegraphics[width=0.9\linewidth]{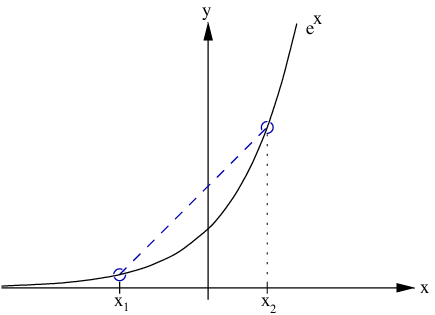}
    \captionsetup{width=.935\linewidth ,justification=raggedright, singlelinecheck=false}
        \captionof{figure}{The plot displays the function $f(x)=\exp(x)$ with two points: $(x_1,\exp(x_1))$ and $\left(x_2,\exp\left(x_2\right)\right)$. A dashed (blue) line connects these points. }\label{fig_convex}
\end{figure}

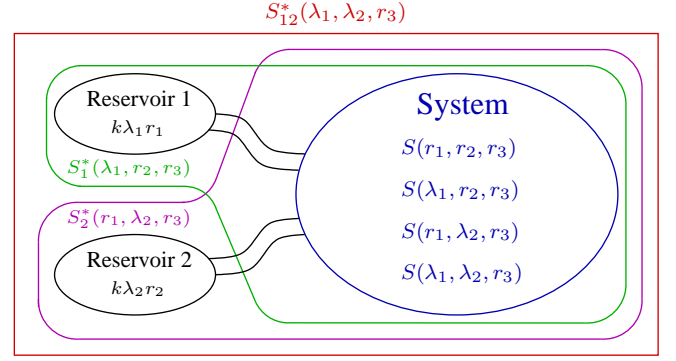
\begin{figure}
\begin{picture}(0,0)%
\includegraphics[width=1\linewidth]{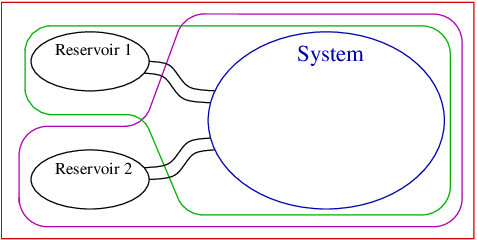}%
\end{picture}%
\setlength{\unitlength}{5180sp}%
\begingroup\makeatletter\ifx\SetFigFont\undefined%
\gdef\SetFigFont#1#2#3#4#5{%
  \reset@font\fontsize{#1}{#2pt}%
  \fontfamily{#3}\fontseries{#4}\fontshape{#5}%
  \selectfont}%
\fi\endgroup%
\begin{picture}(3624,1600)(-1811,-73)
\put(-600,1550){\makebox(0,0)[lb]{\smash{{\SetFigFont{8}{10.8}{\rmdefault}{\mddefault}{\updefault}{\color[rgb]{.82,0,0}$S_{12}^*(\lambda_1,\lambda_2,r_3)$}%
}}}}
\put(50,900){\makebox(0,0)[lb]{\smash{{\SetFigFont{8}{10.8}{\rmdefault}{\mddefault}{\updefault}{\color[rgb]{0,0,.69}$S(r_1,r_2,r_3)$}%
}}}}
\put(50,700){\makebox(0,0)[lb]{\smash{{\SetFigFont{8}{10.8}{\rmdefault}{\mddefault}{\updefault}{\color[rgb]{0,0,.69}$S(\lambda_1,r_2,r_3)$}%
}}}}
\put(50,500){\makebox(0,0)[lb]{\smash{{\SetFigFont{8}{10.8}{\rmdefault}{\mddefault}{\updefault}{\color[rgb]{0,0,.69}$S(r_1,\lambda_2,r_3)$}%
}}}}
\put(50,300){\makebox(0,0)[lb]{\smash{{\SetFigFont{8}{10.8}{\rmdefault}{\mddefault}{\updefault}{\color[rgb]{0,0,.69}$S(\lambda_1,\lambda_2,r_3)$}%
}}}}
\put(-1550,810){\makebox(0,0)[lb]{\smash{{\SetFigFont{7}{10.8}{\rmdefault}{\mddefault}{\updefault}{\color[rgb]{0,.69,0}$S_1^*(\lambda_1,r_2,r_3)$}%
}}}}
%\put(-1550,1380){\makebox(0,0)[lb]{\smash{{\SetFigFont{6}{10.8}{\rmdefault}{\mddefault}{\updefault}{\color[rgb]{0,.69,0}$S_1^*(\lambda_1,r_2,r_3)$}%
%}}}}
\put(-1340,230){\makebox(0,0)[lb]{\smash{{\SetFigFont{7}{10.8}{\rmdefault}{\mddefault}{\updefault}{\color[rgb]{0,0,0}$k\lambda_2 r_2$}%
}}}}
\put(-1340,1000){\makebox(0,0)[lb]{\smash{{\SetFigFont{7}{10.8}{\rmdefault}{\mddefault}{\updefault}{\color[rgb]{0,0,0}$k\lambda_1 r_1$}%
}}}}
\put(-1550,570){\makebox(0,0)[lb]{\smash{{\SetFigFont{7}{10.8}{\rmdefault}{\mddefault}{\updefault}{\color[rgb]{.69,0,.69}$S_2^*(r_1,\lambda_2,r_3)$}%
}}}}
\end{picture}%
        \captionsetup{width=.935\linewidth ,justification=raggedright}
        \captionof{figure}{This is an illustration of the experimental setup of a system with two reservoirs. The system includes three measurements: $r_1$, $r_2$, and $r_3$ and contains the entropy $S$. The combined entropy of the system and reservoir 1 is denoted as $S_1^{*}$ and the combination with reservoir 2 is denoted as $S_2^{*}$. The total entropy including both reservoirs 1 and 2 is denoted as $S_{12}^{*}$.}\label{fig_SystemReservoir}
\end{figure}

      From an experimental perspective, it can be advantageous to adjust the associated conjugate quantity $\lambda_i$ instead of measuring certain quantities $r_i$. In statistical physics, this means describing the entropy in terms of $\lambda_i$ rather than $r_i$. This approach allows us to describe the function $S(r)$ with its slope $\lambda_i(r)\stackrel{\text{exp}}{=}\frac{1}{k}\frac{\partial S(r)}{\partial r_i}$.
      The technical process for achieving this is outlined in fig. \ref{fig_Lt}, there is an example of a quadratic function.
  The diagram demonstrates how to alter the dependence without losing information. To encode the information of the position $x_0$ and the associated value $y_0$ in a function $f^*(s)$, we require a second characteristic of the tangent: the offset of the tangent $d$, which is dependent on the slope $s$.  Geometrically, there are two ways to convert the full information of $f(x)$:
\begin{align}
    \pm f^*(s)= d(x(s)) = y- s \, x= f(x(s)) - s \, x(s) \label{2_Leg}
\end{align}
In eq. \eqref{2_Leg}, we can see that the describing function resulting from using the slope $s$ instead of $x$ is the slope dependence offset $d(s)$. The question that naturally arises is: what are the necessary conditions for this type of transformation to be unique? Fortunately, mathematicians have answered this question: $f^*(s)$ is only a function, with a unique $d$ value for each $s$, if $f$ is convex. To gain an understanding of convex functions, we will demonstrate the concept geometrically using the function $f(x) = \exp(x)$ in fig. \ref{fig_convex}.
    There, it is shown that the line $y(x)$ between the two points $(x_1,\exp(x_1))$ and $(x_2,\exp(x_2))$ is parameterized as: 
    \begin{equation}
         y(t)= \exp(x_2)t +\exp(x_1)(1-t)
    \end{equation}
    with $t=\frac{x-x_1}{x_2-x_1}$. Therefore, for $t \in [0,1]$ the line is restricted between the points $(x_1,\exp(x_1))$ and $(x_2,\exp(x_2))$ and it is always greater than $f(x)=\exp(x)$ for $ x_1\leq x \leq x_2 $. This property defines a convex function. Convexity allows us to properly define the transformation in eq. \eqref{2_Leg}: 
\begin{definition}{Legendre transformation}{}
    Let $f:\mathbb{R}^J \rightarrow \mathbb{R}$ be a convex function, i.e. $\forall t \in [0,1]: \forall r,s \in \mathbb{R}^J:$
    $$   f(t\cdot r+(1-t)\cdot s)\leq tf(r)+(1-t)f(s) $$
    Then the Legendre transformation (LT) of $f$ with $r_i\rightarrow \lambda_i $, written as 
    $f_i^*(r_1,\dots,\lambda_i,\dots,r_j)$ is defined as
    \begin{equation}
        f_i^*(r_1,\dots,\lambda_i,\dots,r_j) \coloneqq\sup_{r_i \in \mathbb{R}}(r_i \lambda_i -f(r)) \label{supremum}
    \end{equation}
\end{definition}
    The LT possesses significant mathematical properties, which we will reiterate by utilizing the results on our entropy. These properties are crucial:
\begin{itemize}
    \item The function $f^*$ is once again convex. This property is encountered in thermodynamics and classical mechanics. When starting from a convex potential, the LT inherits this property. In application, if there is an energy minimum, the transformed energy is also realized as a minimum.
    \item $f$ and $f^*$ are linked symmetrically:
    \begin{equation}  
        \forall r,\lambda \in \mathbb{R}: \, f(r)+f^*(\lambda) \geq r \lambda  \label{Euler_lag}
    \end{equation}
    The equality in eq. \eqref{Euler_lag} holds when $r$ maximises the supremum in eq. \eqref{supremum}, we will denote this by $r(\lambda)$. This property will be crucial in the discussion of thermodynamics.\\
    \item The LT is an \textbf{involution}, therefore 
    \begin{equation}
        (f^*)^*(r)=\sup_{\lambda \in \mathbb{R}}(r \lambda - f^*(\lambda))=f(r)
    \end{equation}
    The complete information has been preserved as desired.
    \item If $f$ is differentiable we can calculate the maximum of $r\lambda-f(r)$ and find $r(\lambda)$. This results in the following equation:
    \begin{align}
     \lambda=f^\prime(r(\lambda))   
    \end{align}
    The unique solution of this equation is $r(\lambda)$, so we get for the LT of $f$
    \begin{equation}
        f^*(\lambda)=r(\lambda) f^\prime(r(\lambda)) -f(r(\lambda)) \label{LT_practical}
    \end{equation}
\end{itemize}
Although the LT is typically defined for convex functions, this concept can also be extended to concave functions such as entropy. In our case, the entropy $S(r)$ is differentiable, leading to the following definition 
\begin{definition}{Concave}{}
    A function $g$ is considered concave if its negative, $-g$, is convex. The LT of a concave function $g$ is defined as 
    \begin{equation}
        g^*(\lambda) \coloneqq-r(\lambda) g^\prime(r(\lambda)) +g(r(\lambda)) \label{LT_concave}
    \end{equation}
\end{definition}
   It is important to note that while the entropy $S(r)$ is concave, $S^*(\lambda)$ is convex and $(S^*)^*(r)=S(r)$ is concave again. This highlights the difference between whether the function is concave or convex, which must also be considered in its application: In mechanics, the kinetic energy is a convex function,  To switch between the Lagrange and Hamiltonian formalism, we use eq. \eqref{LT_practical}. However, in statistical physics, we deal with the \textbf{concave} entropy. By applying eq. \eqref{LT_concave} to our concave entropy $S$, we have already calculated the necessary component in eq. \eqref{exp_mod_comb}: 
    $$ \frac{\partial S(r)}{\partial r_i}=k \lambda_i $$
    As previously mentioned, $r_i$ and $k\lambda_i$ have a special relationship as entropy-conjugated variables. This relationship also reappears in the LT. Therefore it is worth introducing the notation for entropy-conjugated variables:
    $$ r_i \stackrel{S}{\Leftrightarrow} k \lambda_i$$
       To clarify the concept of LT and entropy-conjugated variables, let us demonstrate using the most relevant experimental example: \\
\textbf{LT of $S$ concerning only the energy:}\\
    Assuming $J=1$ and the only observable is $\Op{R}_1=\Op{H}$, the Hamiltonian with energy $E$ as the measurement value $r_1=E$. Then the entropy $S$ can be calculated using eq. \eqref{S_final}    \begin{equation}
        S(E)=-k \Omega(\lambda(E)) +k \lambda(E) E   
    \end{equation}
    Suppose the measured value $E$ is adjusted using a reservoir. A reservoir refers to a system larger than the one considered and acts as a buffer, thus operating as a sink or source of entropy. We want to include the exchange with the reservoir in the entropy calculation, resulting in the \textbf{combined} entropy denoted as $S^*$. To calculate this, we use the LT and require the entropy-conjugated variable of $E$, which is defined from a measurement as
    \begin{equation}
        \beta(E)\equiv \lambda(E)\stackrel{\text{exp}}{=}\frac{1}{k}\frac{\partial S(E)}{\partial E}  \label{def_beta}
    \end{equation}
    with $\beta$ as the coldness, which is related to the temperature $T$ via
        \begin{equation}
        T=\frac{1}{k \beta }
    \end{equation}
    assuming $k$ is fixed to the Boltzmann-Constant. The \textbf{coldness} $\beta$ parameterises the entropy exchange between the system we want to study and the reservoir. To calculate $S^*$, we can use the experimental knowledge $E \stackrel{S}{\Leftrightarrow} k \beta$ and eq. \eqref{LT_concave}:
    \begin{subequations}
    \begin{equation}
        S^*(\beta)= S(E(\beta)) - k \beta E(\beta) \label{S*_1}
    \end{equation}
    The first part of eq. \eqref{S*_1} represents the entropy of the original system, now depending on $\beta$, while the second part represents the exchanged entropy with the reservoir. By inserting eq. \eqref{S_final} into eq. \eqref{S*_1} we find that
    \begin{equation}
        S^*(\beta)= -k \Omega(\underbrace{\beta(E(\beta))}_{\beta}) \overbrace{+k \underbrace{\beta(E(\beta))}_{\beta} E(\beta) - k \beta E(\beta)}^{0}
    \end{equation}
    which can be simplified to
    \begin{equation}
        S^*(\beta)= -k \Omega(\beta) \label{S*_full}
    \end{equation}
    \end{subequations}
    As shown in eq. \eqref{S*_full}, the fully Legendre transformed $S^*$ can be calculated directly by knowing $\Omega$ from eq. \eqref{Omeiga_fertig}. This means that the entropy including all reservoirs is purely determined by the coupling parameter namely the coldness $\beta$. This concept can be easily extended to a system with any number of measurements including their respective reservoirs:
    \begin{equation}
         S^*(\lambda)= -k \Omega(\lambda) \label{S*_generel}
    \end{equation}
    In this discussed case, with only one reservoir having the coupling $\beta$, the combined entropy $S^*(\beta)$ is referred to as the \textbf{free entropy} and is linked to the \textbf{free energy} $F$ through
    \begin{equation}
        F(\beta) \coloneqq  -\frac{S^*(\beta)}{k\beta}=\frac{1}{\beta}\Omega(\beta)=-\frac{1}{\beta}\ln\Bigg(\Tr\bigg( \exp(-\beta\Op{H})\bigg)\Bigg) \label{F_stat}
    \end{equation}
   In cases where $J>1$, there are various options for LTs depending on the experimental setup, and the quantities of interest, as shown in fig. \ref{fig_SystemReservoir}.
%Abbildung fuer reservoirs
This figure allows us to provide a physical interpretation for the mathematically geometric LT in the general case where $J>1$. The LT expands our entropy to include the selected reservoir: $S_1^{*}(\lambda_1,r_2,r_3)$ represent the entropy from our system, including reservoir 1, through the coupling $k\lambda_1$:
   \begin{subequations}
   \begin{align}
       S_1^{*}(\lambda_1,r_2,r_3)=S(\lambda_1,r_2,r_3)-k\lambda_1r_1(\lambda_1,r_2,r_3) \label{LT_2res_1}
   \end{align}
   The first part of eq. \eqref{LT_2res_1} represents the entropy of the original system, and the second represents the exchanged entropy with reservoir 1.  The repetition of the same LT of $S_1^{*}(\lambda_1,r_2,r_3)$ eq. \eqref{LT_2res_1} results in the removal of reservoir 1:
   \begin{align}
     \hspace{0pt}  \left(S_1^{*}\right)_1^{*}(r_1,r_2,r_3) &=S_1^{*}(r_1,r_2,r_3)+k\lambda_1(r_1,r_2,r_3)r_1\notag \\ 
     =S(r_1,r_2,r_3)&\underbrace{-k\lambda_1(r_1,r_2,r_3)r_1+k\lambda_1(r_1,r_2,r_3)r_1}_{=0} \label{LT_2res_11}
   \end{align}
  %From the second LT eq. \eqref{LT_2res_11} we observe that the property $S_1^{*}(r_1,r_2)$ is the entropy of our system \textbf{excluding the knowledge} of the measurement of $\Op{R}_1$. 
  An alternative approach is to incorporate reservoir 2 by means of LT eq. \eqref{LT_2res_1}, using the coupling $k\lambda_2$:
      \begin{align}
       S_{12}^{*}(\lambda_1,\lambda_2,r_3)=S_1^{*}(\lambda_1,\lambda_2,r_3)-k\lambda_2r_2(\lambda_1,\lambda_2,r_3) \label{LT_2res_12}
   \end{align}
   The first part of eq. \eqref{LT_2res_12} describes the entropy of the system including the coupling to reservoir 1 and the second part the entropy exchanged with the reservoir 2. To exclude reservoir 1 or/and 2, LT can be applied again. 
   %Doing that one can determine $S^{**}(\lambda_1,r_2)$, $S^{**}(r_1,\lambda_2)$ and $S^{**}(r_1,r_2)$ with similar interpretation like we did in eq. \eqref{LT_2res_11}.
   The same discussion as in fig. \ref{fig_SystemReservoir} can also be done by starting LT to include reservoir 2, leading to the same result. 
      \end{subequations}
   It is important to note that all properties in fig. \ref{fig_SystemReservoir} can be calculated if $S(r_1,r_2,r_3)$, $S_1^{*}(\lambda_1,r_2,r_3)$, $S_2^{*}(r_1,\lambda_2,r_3)$ or $S_{12}^{*}(\lambda_1,\lambda_2,r_3)$ is known. It should be noted that although they have different experimental meanings, they are connected via the LT.  \\ 
   In practice, dealing with such systems involves first calculating the combined entropy eq. \eqref{S*_generel} including all reservoirs. To obtain the desired information, one can then start Legendre transforming $S^*(\lambda)$. Although $S^*(\lambda)$ is convex by construction, it is necessary to use the LT eq. \eqref{LT_concave} because the original entropy $S$ is concave. This formalism allows for the calculation of the entropy of the original system and the exchange entropy with the associated reservoir. %Such setups can be generalized arbitrarily. 
   \section{Full Formalisation of the Experimental Setup}\label{sec:form}
  Experimental setups can become complex. To control this complexity, we formalise the concept shown in fig. \ref{fig_SystemReservoir}.
  To determine the \textbf{number} and \textbf{type} of observables considered in the experiment, an \textbf{ensemble} is introduced:
\begin{definition}{Statistical Ensembles}{}
    A statistical ensemble is a list of observables $\big(\{\Op{R}_j\}_{j=1}^J \big)$, denoted as ($\Op{R}_1, \Op{R}_2,\dots,\Op{R}_J$).
\end{definition}
    To represent such an ensemble we have already utilised
    $$ 
         C_r \coloneqq r =(r_1,r_2,\dots, r_J)
    $$
    $$ 
        C_\lambda \coloneqq \lambda= (\lambda_1,\lambda_2,\dots,\lambda_J)
    $$
    giving us the familiar entropy that depends on $r$ or $\lambda$
    $$ S(C_r)=S(r) \text{ or } S(C_\lambda)=S(\lambda) $$
    To generalize this concept, one must choose either $r_i$ or $k\lambda_i$ for each reservoir presented in the experimental setup. It is not possible to have a dependency of both since they are entropy-conjugated $r_i \stackrel{S}{\Leftrightarrow} k \lambda_i$. This choice between $r_i$ and $k\lambda_i$ is called a representation of the statistical ensemble ($\Op{R}_1, \Op{R}_2,\dots, \Op{R}_J$):
\begin{definition}{Representation of a Statistical Ensembles}{}
    Consider a statistical ensemble denoted by ($\Op{R}_1, \Op{R}_2,\dots, \Op{R}_J$), where the exclusive choice between $r_i$ and its entropy-conjugated quantity $k\lambda_i$ is made for every present reservoir $k\lambda_i r_i$. If no reservoir is present, $r_i$ is chosen, resulting in a \textbf{representation} $C$ of the statistical ensemble.
\end{definition}
    Examples of statistical ensembles and their representations are:
\hspace*{-15pt}\begin{tabular}{c l c l c l c l} 
    & & & & & & \\
    $(\Op{H})$& $\Rightarrow$ &$(E)$&$\Leftrightarrow$& $(\beta)$& & & \\ [0.5ex]
    $(\Op{N})$& $\Rightarrow$ &$(\expval{\Op{N}})$&$\Leftrightarrow$& $(\beta \mu)$& & & \\ [0.5ex]
    $(\Op{H},\Op{N})$& $\Rightarrow$ &$(E,\expval{\Op{N}})$&$\Leftrightarrow$& $(\beta,\expval{\Op{N}})$&$\Leftrightarrow$& $(E, \beta \mu)$ &$\Leftrightarrow (\beta,\beta\mu)$\\
    & & & & & & \\
\end{tabular}\\
Consider an ensemble $(\Op{R}_1,\Op{R}_2,\Op{R}_3,\Op{R}_4)$ with two reservoirs having coupling parameters $k\lambda_2$ and $k\lambda_3$. The representations $C$  of this ensemble are as follows:
\begin{align*}
    C_r&=(r_1,r_2,r_3,r_4) \\
    C_1&=(r_1,r_2,\lambda_3,r_4) \\
    C_2&=(r_1,\lambda_2,r_3,r_4)  \\
    C_3&=(r_1,\lambda_2,\lambda_3,r_4)
\end{align*}
As outlined in the definition of representations $C$,  $r_1$ and $r_4$ must be selected, as there is no coupling to their respective reservoirs. \\
For any statistical ensemble and a given representation $C$, it is possible to calculate the entropy of the original system as
\begin{equation}
     S(C)=-k \Omega(C) + k \sum_{\lambda_i \in C} \lambda_i r_i(C) +k \sum_{r_i \in C}\lambda_i(C) r_i  \label{S_C}
\end{equation}
where
\begin{equation}
    \Omega(C)=-\ln\!\left(\!\Tr\!\left(\!\exp\!\!\Bigg(\!\!-\!\!\! \sum_{\lambda_i \in C} \lambda_i \Op{R}_i - \sum_{r_i \in C}\lambda_i(C) \Op{R}_i \!\!\Bigg)\!\!\right)\!\!\right)
 \label{Omega_C}
\end{equation}
As previously stated, this is only a portion of the total entropy of the system. 
 In order to calculate a combined entropy $S_{C'}^*(C)$, one must first select a representation $C$ and a list of coupling parameters $\lambda_i$ for the chosen reservoirs, which are encoded in the representation $C'$:
\begin{align}
    S_{C'}^*(C)=-k \Omega(C) + k \sum_{\substack{\lambda_i \in C \\\  \lambda_i \notin C'}} \lambda_i r_i(C) +k \sum_{\substack{r_i \in C \\ \ r_i \in C'}}\lambda_i(C) r_i  \label{S_C2}
\end{align}
    \textbf{Statistical potentials} can be derived from eq. \eqref{S_C2}, by setting $C=C'$. 
\begin{definition}{Statistical Potentials}{}
    A statistical potential is the LT, according to $C'$, of $S(r)$ with the representation $C=C'$ of the statistical ensemble  ($\Op{R}_1, \Op{R}_2,\dots, \Op{R}_J$):
    \begin{equation}
        S_C^*(C)=-k \Omega(C)  +k \sum_{r_i \in C}\lambda_i(C) r_i \label{stat_pot}
    \end{equation}
\end{definition}
     Potentials and their associated representations are described using \textbf{natural variables} ($C=C'$). Although eq. \eqref{S_C2} appears more general, only the statistical potentials eq. \eqref{stat_pot} contain the full information on the experimental setup. It is only from these statistical potentials that every other entropy eq. \eqref{S_C2} can be calculated.
  \\Finally, there is one question left to answer: 'What does this have to do with thermodynamics?' 
\section{From Statistical Physics to Thermodynamics} \label{sec: Thermo}
    To derive thermodynamics (TD) from statistical physics, it is shown how to get the axioms of thermodynamics \cite{neumaier2011classical}.
    In TD, the primary quantity of interest is energy $E$. To provide a historical perspective, the focus must shift from entropy to energy. To accomplish this, we consider that in thermodynamics, the energy observable $\Op{H}$ is always present. Without loss of generality, we select $\Op{R}_1=\Op{H}$, yielding $r_1=E$. Its entropy-conjugated quantity $E \stackrel{S}{\Leftrightarrow} k \beta$ corresponds to the coldness $\lambda_1=\beta$. By inserting this information into eq. \eqref{S_final}, we find 
\begin{subequations}
        \begin{equation}
            S(E,r)=-k \Omega(E,r) + k\beta(E,r) E + k \sum_{j=2}^J \lambda_j(E,r) r_j \label{S_therm_beg}
        \end{equation}
    where $r=(r_2,\dots,r_j)$. To express the entropy-conjugated quantities through \textbf{energy-conjugated} ones, we need to alter the point of view from entropy $S(E,r)$ to energy $E(S,r)$. This requires using the chain rule on the entropy-conjugated quantities:
    \begin{equation}
         k \lambda_j=\frac{\partial S}{\partial r_j}=\frac{\partial S}{\partial E}\frac{\partial E}{\partial r_j} = k\beta \frac{\partial E}{\partial r_j} \label{chain}
    \end{equation}
	Using the chain rule eq. \eqref{chain}, we can define the energy-conjugated variables as 
\begin{equation}
     r_i \stackrel{E}{\Leftrightarrow} k\Lambda_i
\end{equation}
with
\begin{equation}
    \Lambda_i \coloneqq\frac{\partial E}{\partial r_i} \label{Lamda}
\end{equation}
Upon examining eq. \eqref{chain} and eq. \eqref{Lamda}, it becomes apparent that the variables conjugated to entropy and energy are linked through the variable $\beta$, specifically
\begin{equation}
    \lambda_i=\beta\Lambda_i \label{translat}
\end{equation}
Using identity eq. \eqref{translat}, we can express entropy eq. \eqref{S_therm_beg} and eq. \eqref{Omeiga_fertig} in terms of energy-conjugated variables: 
    \begin{equation}
        \begin{split}
             S(E,r)=&-k \Omega(E,r)+k\beta(E,r) E \\
                    &+ k \beta(E,r)\sum_{j=2}^J \Lambda_j(E,r)\,  r_j \label{S_energy_con}
        \end{split}
    \end{equation}
\begin{equation}
    \hspace*{-5pt}\Omega(E,r)=-\ln\!\left(\!\Tr\!\left(\!\exp\!\!\Bigg(\!\!-\beta(E,r)\!\bigg(\Op{H} +\!\!\sum_{j=2}^J  \Lambda_j(E,r) \Op{R}_j\bigg)\!\!\Bigg)\!\!\right)\!\!\right)
\end{equation}
%\begin{equation}
%    \Omega(E,r) \coloneqq-\ln\left(\Tr\left(\exp\left(-\beta(E,r)\left(\Op{H} +\sum_{j=2}^J  \Lambda_j(E,r) \Op{R}_j\right)\right)\right)\right)
%\end{equation}
  \end{subequations}
Some examples of conjugated variables are
\begin{center}
\begin{tabular}{c c c c c c c }
    %& & & & & & \\
    $E \stackrel{S}{\Leftrightarrow} k \beta$, & & $S \stackrel{E}{\Leftrightarrow} \frac{1}{k\beta}$, & &  $N \stackrel{E}{\Leftrightarrow} \mu$,& &$X \stackrel{E}{\Leftrightarrow} \frac{\partial E}{\partial X}$\\
    %& & & & & & \\
\end{tabular}
\end{center}
%Vertical
%\begin{itemize}
%    \item $E \stackrel{S}{\Leftrightarrow} k \beta$
%    \item $S \stackrel{E}{\Leftrightarrow} \frac{1}{k\beta}$
%    \item $N \stackrel{E}{\Leftrightarrow} \mu$
%    \item $X \stackrel{E}{\Leftrightarrow} \frac{\partial E}{\partial X}$
%\end{itemize}
    From a statistical ensemble, which includes the Hamiltonian ($\Op{H},\{\Op{R}_j\}_{j=2}^J$), we can construct a TD ensemble. This TD ensemble differs only in the representation. One must choose between $r_i$ and $\Lambda_i$. In TD, we mainly deal with three ensembles: the micro-canonical, canonical, and grand canonical ensembles. In the microcanonical ensemble, the probabilities of the state are equally distributed as there are no measurements. In the canonical ensemble, the only observable is $\Op{H}$, resulting in the Boltzmann distribution. In the grand canonical ensemble, there are two observables: the Hamiltonian $\Op{H}$ and the number of particles $\Op{N}$.
    A Hamiltonian $\Op{H}$ that is \textbf{linear} in $\Op{N}$ leads to the Bose-Einstein distribution for bosons and the Fermi-Dirac distribution for fermions. When calculating these ensembles, TD potentials are also introduced. By applying the chain rule to the same conversion procedure, one can find 
    \begin{equation}
       U_{C}^*(C) \coloneqq-\frac{ S_C^*(C)}{k\beta(C)} \label{TD}
    \end{equation}
    The TD potential in eq. \eqref{TD} is represented by $U_{C}^*(C)$. An example of such a potential has already been encountered in eq. \eqref{F_stat}, where the potential was referred to as the free energy. If no reservoirs are present, meaning the only representation is $C=C_r$, the TD potential is then called the inner energy $U$, which is defined as 
   \begin{equation}
       U(r) \coloneqq U_{C_r}^*(C_r).
    \end{equation}
    The inner energy serves as the starting potential for every thermodynamic analysis. All relevant quantities can be extracted from the TD potential by differentiation. If the coldness $\beta$ is part of the representation $C$ the entropy can be recovered, by using eqs. \eqref{TD} and \eqref{stat_pot}  
    \begin{subequations}
    \begin{equation}
        \pdv{U_{C}^*(C)}{\beta}=-\frac{1}{\beta^2}\Omega(C)+\frac{1}{\beta}\pdv{\Omega(C)}{\beta}-\sum\limits_{r_i \in C}\pdv{\Lambda_i(C)}{\beta}r_i \label{U_deriv}
    \end{equation}
    where
    \begin{equation}
        \Omega(C)=-\ln\!\left(\!\Tr\!\left(\!\exp\!\!\Bigg(\!\!-\!\!\! \beta \bigg(\Op{H}+\sum_{\Lambda_i \in C} \Lambda_i \Op{R}_i + \sum_{r_i \in C}\Lambda_i(C) \Op{R}_i\bigg) \!\!\Bigg)\!\!\right)\!\!\right)
        \label{Omega_Ct}
    \end{equation} 
      The second term in eq. \eqref{U_deriv} is calculated as 
      \begin{align}
          \pdv{\Omega(C)}{\beta}=&E(C)+ \sum_{\Lambda_i \in C} \Lambda_i r_i(C) + \sum_{r_i \in C}\Lambda_i(C) r_i \notag\\
          &+\beta \sum_{r_i \in C}\pdv{\Lambda_i(C)}{\beta} r_i \label{U_deriv2}
      \end{align}
      Inserting eq. \eqref{U_deriv2} into eq. \eqref{U_deriv} one finds
      \begin{align}
           \pdv{U_{C}^*(C)}{\beta}=&-\frac{1}{\beta^2}\Omega(C)+\frac{1}{\beta}E(C)+\frac{1}{\beta} \sum_{\Lambda_i \in C} \Lambda_i r_i(C)\notag\\
           &+\frac{1}{\beta} \sum_{r_i \in C}\Lambda_i(C) r_i  \label{U_deriv3}
      \end{align}
    By using eq. \eqref{S_C} and eq. \eqref{translat}, eq. \eqref{U_deriv3} can be simplified to obtain the final result
    \begin{equation}
        \pdv{U_{C}^*(C)}{\beta}=\frac{1}{k\beta^2}S(C) \label{U_deriv_final}
    \end{equation}
        \end{subequations}
    Eq. \eqref{U_deriv_final} shows that starting from the TD potential allows for the reconstruction of entropy, which is a common practice in thermodynamics. The properties of the system are calculated based on the potential. In TD, temperature $T$ is typically used instead of coldness $\beta$. By applying the chain rule to eq. \eqref{U_deriv_final} with respect to $\beta$, we arrive at a fundamental result in TD:
      \begin{equation}
       \pdv{U_{C}^*(C)}{T}=-S(C) \label{U_deriv_final2}
    \end{equation}
    In the context of \textbf{phenomenological TD}, temperature is considered a fundamental property. By measuring the temperature-dependent potential, one can directly observe the entropy as the slope of the potential eq. \eqref{U_deriv_final2}. \\
    To conclude the discussion on TD, we will establish the connection to the axioms of \textbf{phenomenological TD}. 
    These axioms include the Euler inequality and the state function, which we will prove using statistical physics
\begin{axiom}{State function}{State function}
There exists a convex function $\Delta(\beta,\Lambda)$ dependent solely on energy-conjugate variables $\beta$ and $\Lambda$, termed as the state function, which is monotonically increasing in $\beta$ and monotonically increasing (decreasing) in $\Lambda_i$ if $r_i < 0\, (> 0)$. The energy-conjugate quantities must satisfy the equilibrium condition
\begin{equation}
 \Delta(\beta,\Lambda)=0 \label{equi}
\end{equation}
The set of $\beta$ and $\Lambda$ values satisfying the equilibrium condition eq. \eqref{equi} is referred to as the state space. 
\end{axiom}
Phenomenological thermodynamics begins with the equation of state \eqref{equi}. This function is constructed using experimental data, more fundamental theories such as statistical physics, or models. In statistical physics, the state function is represented by
\begin{equation}
     \Delta(\beta,\Lambda) \coloneqq -\frac{\Omega(\beta,\Lambda)}{\beta}
     \label{Stat_thermo}
\end{equation}
Eq. \eqref{F_stat} previously encountered a state function of this kind.\\
The proof of convexity, which satisfies the axiom from above, is given in Appendix \ref{A.Delta}.
The second axiom of TD discusses the Euler inequality.
\begin{axiom}{Euler inequality}{Euler inequality}
    The energy $E$ satisfies the Euler inequality
    \begin{equation}
    E \geq TS- \sum_{j=2}^J r_i \Lambda_i \label{euler}
    \end{equation}
    for all $T$ and $\Lambda$ in the state space. In equilibrium equality holds.
\end{axiom}
\begin{subequations}
From eq. \eqref{Euler_lag} and the concavity of $S$, it can be observed that 
\begin{equation}
    S_{C_\lambda}^*(\beta,\lambda)-S(E,r)\geq  - k \beta E -  k \sum_{j=2}^J r_j \lambda_j   
\end{equation}
Using eq. \eqref{S*_generel} 
\begin{equation}
    k \beta E \geq  k\Omega(\beta,\Lambda) + S(E,r)  -  k \sum_{j=2}^J r_j \lambda_j   
\end{equation}
and by dividing both sides by $k\beta$ and utilizing eq. \eqref{translat} we obtain
\begin{equation}
     E \geq  \frac{\Omega(\beta,\Lambda)}{\beta} + \frac{S(E,r)}{k \beta }  -  \sum_{j=2}^J r_j \Lambda_j   
\end{equation}
Using eq. \eqref{Stat_thermo} one arrives at
\begin{equation}
     E \geq  -\Delta(\beta,\Lambda) + \frac{S(E,r)}{k \beta }   -  \sum_{j=2}^J r_j \Lambda_j   
\end{equation}
Finally, by restricting $\beta$ and $\Lambda$ to the state space eq. \eqref{equi} and expressing $\beta$ in terms of $T$ we arrive at the Euler inequality:
\begin{equation}
     E \geq T S(E,r)  -  \sum_{j=2}^J r_j \Lambda_j 
\end{equation}
\end{subequations}
Eqs. \eqref{equi}  and \eqref{euler} show that TD focuses on equilibrium conditions, while statistical physics encompasses both equilibrium and non-equilibrium physics. However, this generality is accompanied by the disadvantage of not utilising problem-specific properties. In the context of specific problems, the choice between the methods of phenomenological TD and statistical physics may be made.
\section{Conclusion}
This paper serves as a pedestrian guide to understanding the three fundamental properties of entropy, namely, the additive property of independent subsystems, the maximisation of uncertainty in unknown systems, and the absence of entropy in known systems. It elucidates these properties through illustrative examples. This work demonstrates how entropy can be determined from experimental data by making the physical intuition mathematically precise, leading to stability conditions and Maxwell relations. Furthermore, the paper elucidates the interplay between theory and experiment. It illustrates how experimental manipulation via reservoirs influences theory via the Legendre transformation, which was explored geometrically, mathematically and physically. This paper provides a precise definition of ensembles with their representation, highlighting the different entropy functions. Additionally, it offers a definition of potentials with their associated natural variables and insights into how they can be used effectively in statistical physics. Finally, the paper concludes by deriving the theoretical frameworks of phenomenological thermodynamics, which consist of the state function, their convexity and the Euler inequality, through statistical physics. 
\begin{acknowledgments}
We express our gratitude to Thomas Winkler and Dr. Kurt Hingerl for the stimulating discussions and critical reading of the manuscript.
\end{acknowledgments}
 \appendix
 \section{Entropy's of Independent Subsystems are Additive}\label{A.S_add}
Assuming that the state $\Op{\rho}_{\text{AB}}$ comprises of two \textbf{independent} subsystems $\Op{\rho}_{\text A}$ and $\Op{\rho}_{\text B}$:
  \begin{equation}
      \Op{\rho}_{\text{AB}}=\Op{\rho}_{\text A} \otimes \Op{\rho}_{\text B} \label{ind_sub}
  \end{equation}
  The entropy of the entire system can be calculated using
\begin{equation}
    S[\Op{\rho}_{\text{AB}}]=-k\Tr_\text{AB} \left( \Op{\rho}_{\text{AB}}\ln\left(\Op{\rho}_{\text{AB}} \right)\right)
\end{equation}
  By inserting the two independent subsystems eq. \eqref{ind_sub} one can arrive at
  \begin{equation}
      \Tr_\text{AB} \left( \Op{\rho}_{\text{AB}}\ln\left(\Op{\rho}_{\text{AB}}\right) \right)=\Tr_\text{AB} \left( \Op{\rho}_{\text A} \otimes \Op{\rho}_{\text B}\ln\left(\Op{\rho}_{\text A} \otimes \Op{\rho}_{\text B}\right) \right)
  \end{equation}
  As systems A and B are \textbf{independent}, it is possible to decompose the tensor product into a sum of logarithms:
  \begin{equation}
      \ln\left(\Op{\rho}_{\text A} \otimes \Op{\rho}_{\text B}\right) =\ln\left(\Op{\rho}_{\text A}\right) +\ln\left(\Op{\rho}_{\text B}\right) 
  \end{equation}
and find 
\begin{align}
  &\Tr_\text{AB} \left( \Op{\rho}_{\text A} \otimes \Op{\rho}_{\text B}\ln\left(\Op{\rho}_{\text A} \otimes \Op{\rho}_{\text B}\right) \right)=\notag \\ &\Tr_\text{AB} \left( \Op{\rho}_{\text A} \otimes \Op{\rho}_{\text B}\left(\ln\left(\Op{\rho}_{\text A}\right)+\ln \left( \Op{\rho}_{\text B}\right) \right)  \right) \label{ln_split}
\end{align}
  By applying the definition of evaluating the trace $\Tr_{\text{AB}}$ as
  \begin{equation}
      \Tr_{\text{AB}}(\Op{\rho}) \coloneqq\Tr_{\text A}\left(\Tr_{\text B}\left(\Op{\rho}\right) \right)=\Tr_{\text B}\left(\Tr_{\text A}\left(\Op{\rho}\right) \right)
  \end{equation}
   eq. \eqref{ln_split} can be simplified to
  \begin{align}
       \Tr_\text{A} \left( \Op{\rho}_{\text A} \ln\left(\Op{\rho}_{\text A}\right)\right)+\Tr_{\text B}\left(\Op{\rho}_{\text B}\ln \left( \Op{\rho}_{\text B}\right) \right) 
  \end{align}
  These terms represent the entropy of the two subsystems. 
  \begin{equation}
      -k\Tr_\text{A} \left( \Op{\rho}_{\text A} \ln\left(\Op{\rho}_{\text A}\right)\right)-k\Tr_{\text B}\left(\Op{\rho}_{\text B}\ln \left( \Op{\rho}_{\text B}\right) \right) =S[\Op{\rho}_{\text A}]+S[\Op{\rho}_{\text B}]
  \end{equation}
  We have proven that the entropy of \textbf{independent} subsystems satisfy
  \begin{align}
      S[\Op{\rho}_{\text {AB}}]=S[\Op{\rho}_{\text A}]+S[\Op{\rho}_{\text B}]
  \end{align}
  as desired. 
 \section{$\Delta$ is Convex} \label{A.Delta}
 To prove $\Delta$ is convex, it is necessary to demonstrate that the Hessian is positive definite. To obtain the Hessian, we require the second derivatives of $\Delta$. For this, we calculate the first derivatives of $\Delta$
 \begin{align}
     \partial_\beta \Delta&= \frac{1}{\beta^2} \Omega -\frac{1}{\beta} \partial_1 \Omega- \frac{1}{\beta} \sum_{j=2}^J \Lambda_j \partial_j \Omega \notag \\
     \partial_{\Lambda_j} \Delta&=- \partial_j \Omega
 \end{align}
  With this, we can proceed to the second derivatives
  \begin{align}
     \partial^2_\beta \Delta&= -\frac{2}{\beta^3} \Omega + \frac{2}{\beta^2} \partial_1\Omega + \frac{2}{\beta^2} \sum_{j=2}^J \Lambda_j \partial_j \Omega\notag \\
     &- \frac{2}{\beta} \sum_{j=2}^J \Lambda_j \partial_1 \partial_j \Omega -\frac{1}{\beta} \partial_1^2 \Omega \notag \\
     &-\frac{1}{\beta} \sum_{i=2}^J \sum_{j=2}^J \Lambda_i \Lambda_j \partial_i \partial_j \Omega \notag\\
     \partial_\beta \partial_{\Lambda_i} \Delta&=- \partial_1 \partial_i \Omega - \sum_{j=2}^J \Lambda_j \partial_i \partial_j \Omega \notag\\
     \partial_{\Lambda_i} \partial_{\Lambda_j} \Delta&=- \beta \partial_i \partial_j \Omega 
 \end{align}
 To demonstrate positive definiteness, it is necessary to show
 \begin{align}
     \forall a \in \mathbb{R}^J: \sum_{k=1}^J \sum_{l=1}^J a_k a_l\partial_k \partial_l \Delta \geq 0  \label{A.posdev}
 \end{align}
 We need to prove that the following expression is for all $a$ positive
 \begin{align}
     a_1^2 \partial^2_\beta \Delta + 2\sum_{k=2}^J a_1 a_k\partial_\beta \partial_{\Lambda_k} \Delta +\sum_{k=2}^J \sum_{l=2}^J a_k a_l\partial_{\Lambda_k} \partial_{\Lambda_l} \Delta  \label{A.expression}
 \end{align}
 By sorting the terms in eq. \eqref{A.expression}, we obtain the following two contributions expressed in relation to $\Omega$
 \begin{align}
  \frac{2a_1^2}{k\beta^3}\left( -k \Omega + k \beta E + \beta \sum_{j=2}^J \Lambda_j r_j  \right) \label{A.first}
 \end{align}
 and
 \begin{align}
  -\frac{1}{\beta}\bigg( &a_1^2 \partial_1^2 \Omega +2  \sum_{j=2}^J a_1(a_1\Lambda_j+a_j\beta) \partial_1 \partial_j \Omega \notag \\
  &+\sum_{i=2}^J\sum_{j=2}^J  (a_1\Lambda_i+a_i\beta) (a_1\Lambda_j+a_j\beta) \partial_i \partial_j \Omega \bigg) \label{A.second}
 \end{align}
  The contribution of the first term eq. \eqref{A.first} is proportional to $S$
 \begin{align}
      \frac{2a_1^2}{k\beta^3}S(E,\Lambda) \label{A.S}
 \end{align} 
The second term eq. \eqref{A.second} is related to the negative definiteness of the Hessian of $\Omega$ in eq. \eqref{negative_def}, with the specific choice of $b \coloneqq(a_1,\{ a_1\Lambda_j+a_j\beta \}_{j=2}^J  )\in \mathbb{R}^J $:
 \begin{align}
     -\frac{1}{\beta} \underbrace{ \sum_{k=1}^J \sum_{l=1}^J b_k b_l \partial_k \partial_l \Omega }_{ \leq 0}  \label{A.Omega}
 \end{align}
 For $\beta \geq 0$, the two contributions eqs. \eqref{A.S} and \eqref{A.Omega} are positive. Therefore, $\Delta$ is indeed convex. 

\bibliography{Lit}% Produces the bibliography via BibTeX.

\end{document}